\documentclass[preprint]{aastex62}
\usepackage{graphicx}

\shorttitle{Mass and radius determination of NS using ATHENA}
\shortauthors{Majczyna et al.}

\begin{document}

\title[Mass and radius determination of NS using ATHENA]{Precision of mass and radius
determination for neutron star using the ATHENA mission.}

\correspondingauthor{Agnieszka Majczyna}
\email{agnieszka.majczyna@ncbj.gov.pl}

\author{Agnieszka Majczyna}
\affiliation{National Centre for Nuclear Research, ul. Andrzeja So\l{}tana 7, 05-400 Otwock, 
Poland;}
\nocollaboration

\author{Jerzy Madej}
\affiliation{Astronomical Observatory, University of Warsaw, Al. Ujazdowskie 4, 00-478 Warszawa, 
Poland;}
\nocollaboration

\author{Miros\l{}aw Nale\.zyty}
\affiliation{Astronomical Observatory, University of Warsaw, Al. Ujazdowskie 4, 00-478 Warszawa, 
Poland;}
\nocollaboration

\author{Agata R\'o\.za\'nska}
\affiliation{Nicolaus Copernicus Astronomical Centre, Polish Academy of Sciences, ul. Bartycka
18, 00-716 Warszawa, Poland;}
\nocollaboration

\author{Bartosz Be\l{}dycki}
\affiliation{Nicolaus Copernicus Astronomical Centre, Polish Academy of Sciences, ul. Bartycka
18, 00-716 Warszawa, Poland;}
\nocollaboration

\begin{abstract}

In this paper we show that X-ray spectral observations of the ATHENA mission, which is planned to 
launch in 2031, can constrain the equation of state of superdense matter. We use 
our well-constrained continuum fitting method for mass and radius determination of the neutron star. 
Model spectra of the emission from a neutron star were calculated using the atmosphere code ATM24. 
In the next step, those models were fitted to a simulated spectra of the neutron star calculated 
for ATHENA's WFI detector, using the satellite calibration files. To simulate the spectra we 
assumed three different values of effective temperatures, surface gravities and gravitational 
redshifts. There cases are related to the three different neutron star masses and radii. This 
analysis allows us to demonstrate the precision of our method and demonstrate the need for a fast 
detector onboard of ATHENA. A large grid of theoretical spectra was calculated  with various 
parameters and a hydrogen-helium-iron composition of solar proportion. These spectra 
were fitted to the simulated spectrum to estimate the precision of mass and radius determination. 
In each case, we obtained very precise mass and radius values with errors in the range 
3--10\% for mass and in the range 2--8\% for radius within the $1\sigma$ confidence error.
We show here that with the ATHENA WFI detector, such a determination could be used to constrain the 
equation of state of superdense neutron star matter.

\end{abstract}

\keywords{stars: neutron -- mission: ATHENA -- radiative transfer}

\section{INTRODUCTION}

Almost 50 years after confirmation of the existence of the neutron star \citep{hewish68}, the 
equation of state of the matter that comprises these stars is still under discussion.  In neutron 
stars, the density in the center of the star exceeds a few times the nuclear density. Many 
theoretical models of the equation of state (EOS) of superdense matter have been proposed,
\citep[see the extensive review by][]{haensel07}. Models have assumed both normal matter and
matter in exotic states, like condensates of pions or kaons, superfluid or superconductive matter,
or even free quarks. Astronomical observations are the only way to verify the EOS of neutron 
stars, because in Earth laboratories we are unable to reproduce conditions similar to neutron star 
interiors. A very important property of theoretical models is the existence of a maximum mass for
the neutron star and a unique mass -- radius relation for each assumed EOS. There exist 
multiple methods to constrain the EOS with astronomical observations.

Astronomers seek to uncover the heaviest neutron stars \citep{antoniadis13, demorest10}. 
Measurements of the maximum mass allows one to exclude those EOS models that predict a maximum 
mass lower than the observed maximum. Comparison of both masses does not allow one for the unique 
determination of the EOS. There exist methods that allow for simultaneous mass and radius
determination, and consequently, to determine the EOS. One such method is
fitting of the observed spectra  
with model atmospheres, but until the necessary high-quality spectra are obtained, the accuracy of 
mass and radius determination will remain an open question. We expect that such high quality spectra 
can be obtained by the ATHENA detectors, especially that a growing population of 
bursters, currently numbered at 110\footnote{http://burst.sci.monash.edu/sources} 
has been observed by almost every major X-ray satellite \citep[][and references therein]{watts2016,galloway2017}.

ATHENA (Advanced Telescope for High Energy Astrophysics) is an X-ray mission accepted by the ESA 
to address the Hot and Energetic Universe science theme \citep{athena-mission}. The mission will 
be launched in 2031 and placed at the second Sun-Earth Lagrangian point (L2). The planned 
mission lifetime is five years, but the mission is expected to last longer. ATHENA will be equipped 
with two scientific instruments: the X-ray Integral Field Unit \citep[X-IFU;][]{athena-XIFU} and the 
Wide Field Imager \citep[WFI;][]{athena-WFI}. 

As per its name, the WFI has a large  field of view 40'$\times$40', and very high angular 
resolution, 5''. It will observe in the energy range 0.2 -- 15 keV with resolution 170 eV at 7 keV. 
The planned time resolution for this instrument is 80 $\mu$sec. It's scientific goals are related to 
high energy phenomena, and include studying hot baryons in groups and clusters of galaxies, 
accretion processes onto compact objects, and GRBs and other transient objects. The high time 
resolution of WFI in combination with the large effective area of the ATHENA mirrors make this 
detector fast enough to be used for studying neutron stars during bursts. Such conditions are needed 
for mass and radius determination when using the continuum fitting method.

The continuum fitting method was first described by \cite{majczyna05}. They fit PCA/RXTE spectra of 
MXB~1728$-$34 taken during a phase between the bursts. Each spectrum was integrated over 16 s. These 
authors fit numerical models calculated with the ATM21 code to the observed spectra. Furthermore, 
the same models were fitted to the observed spectra  of 4U 1820$-$30 \citep{kusmierek11}. They 
obtained values of mass $M=1.3\pm0.6\,$M$_\odot$ and radius $R=11^{+3}_{-2}\,$km consistent with 
results obtained by other researchers. Errors in the paper by \cite{kusmierek11} are relatively 
large but they could be reduced if some systematic effects known now (e.g. accretion during even 
strong bursts) are included. Continuum fitting method for neutron star mass and radius 
determination can be also used without complicated 
calculations of neutron star atmospheres. Instead, black body emission multiplied by
the color correction factor can be assumed \citep{ozel2009}. Such approach is faster, but it does 
not take into account  that in reality, the overall shape of the emitted spectrum is modified by 
Compton scattering, especially at the hard tail of the spectrum \citep{majczyna05,suleimanov2011}.

In our analysis we used fake spectra, therefore, we made the principal assumption that our 
theoretical models are valid for this ``source". Therefore, we did not widely discuss validation of 
each assumption of our model in context of real sources. But we will note that our theoretical 
spectra could be used to fit
the observed spectra of real sources (see eg. \citet{kusmierek11}). In this paper, we clearly show 
that the data which will be provided by WFI/ATHENA will allow 
us to determine mass and radius, using continuum fitting method, with errors as small as 
3-10\% for mass determination and 2-8\% for radius determination even for relatively dim sources.

\section{THE ATM24 MODEL CODE}

The model atmospheres and theoretical X-ray spectra of hot neutron stars used in this paper 
were computed with the ATM24 code, which is the next version of ATM21 code 
\citep{madej91,majczyna05b} upgraded for its numerical precision. The accuracy of the code has 
been recently demonstrated by \citet{madej17,vincent17}. The ATM24 code calculates the radiative 
transfer equation in a plane-parallel geometry. It takes into account the effect of Compton
scattering on free, relativistic electrons, where initial photon energies can approach the electron
rest mass. We assume  the equation of state of ideal gas being in  local thermodyna\-mical
equilibrium (LTE). Nevertheless, the Compton scattering redistribution functions of X-ray photons 
$\Phi(\nu,\nu')$ are fully non-LTE terms of the radiative transfer equation. 

The equation of transfer was adopted from \citet[][see also \citealp{sampson59}]{pomraning73}. The 
working equation of transfer and the temperature correction procedure were presented 
originally by \cite{madej89,madej91} and used correctly by \cite{madej17,vincent17}. The final 
equation of transfer is written on the monochromatic optical depth scale 
$d\tau_\nu=-(k_\nu+\sigma_\nu)\,\rho\, dz$, and  has a form:
\begin{eqnarray}
 \mu\,\frac{d I_\nu}{d \tau_\nu} & =& 
I_\nu-{\frac{k_\nu}{k_\nu+\sigma_\nu}}B_\nu-\left(1-{\frac{k_\nu}{k_\nu+\sigma_\nu}}
\right)J_\nu +
\left(1-{\frac{k_\nu}{k_\nu+\sigma_\nu}}\right) J_\nu \int\limits_0^\infty\Phi(\nu,\nu') \left(1+
{\frac{c^2}{2h{\nu'}^3}}J_{\nu'}\right) d\nu'+ \nonumber \\
& - & {\frac{k_\nu}{k_\nu+\sigma_\nu}}\left(1+{\frac{c^2}{2h\nu^3}}J_\nu\right)
 \times 
\int\limits_0^\infty\Phi(\nu,\nu')J_{\nu'} {\left({\frac{\nu}{\nu'}}\right)}^3\exp\left[-{\frac{
h(\nu-\nu')}{{\rm k}T}}\right] d \nu',
\end{eqnarray}
where $k_\nu$ and $\sigma_\nu$ denote coefficients of absorption and electron scattering,
respectively. $I_\nu$ is the energy-dependent specific intensity, $J_\nu$ is the mean intensity of 
radiation, and $z$ is the geometrical depth in the considered atmosphere.

We used the angle-averaged redistribution function $\Phi(\nu,\nu')$ and Compton scattering 
cross-section $\sigma(\nu\to\nu',\vec{n}\cdot \vec{n'})$, following the method by 
\citet{guilbert81}, which was corrected for the computational error by \citet{madej91}. The Compton 
redistribution function is related to cross section as defined in \citet{madej89}:
\begin{equation}
\Phi(\nu,\nu')={\frac{1}{\sigma_\nu}}\oint_{\omega'}\frac{d\omega'}{4\pi}\,
\sigma(\nu\to\nu',\vec{n}\cdot\vec{n'}).
\end{equation}

\begin{figure}[!h]
\begin{center}
\includegraphics[scale=0.46]{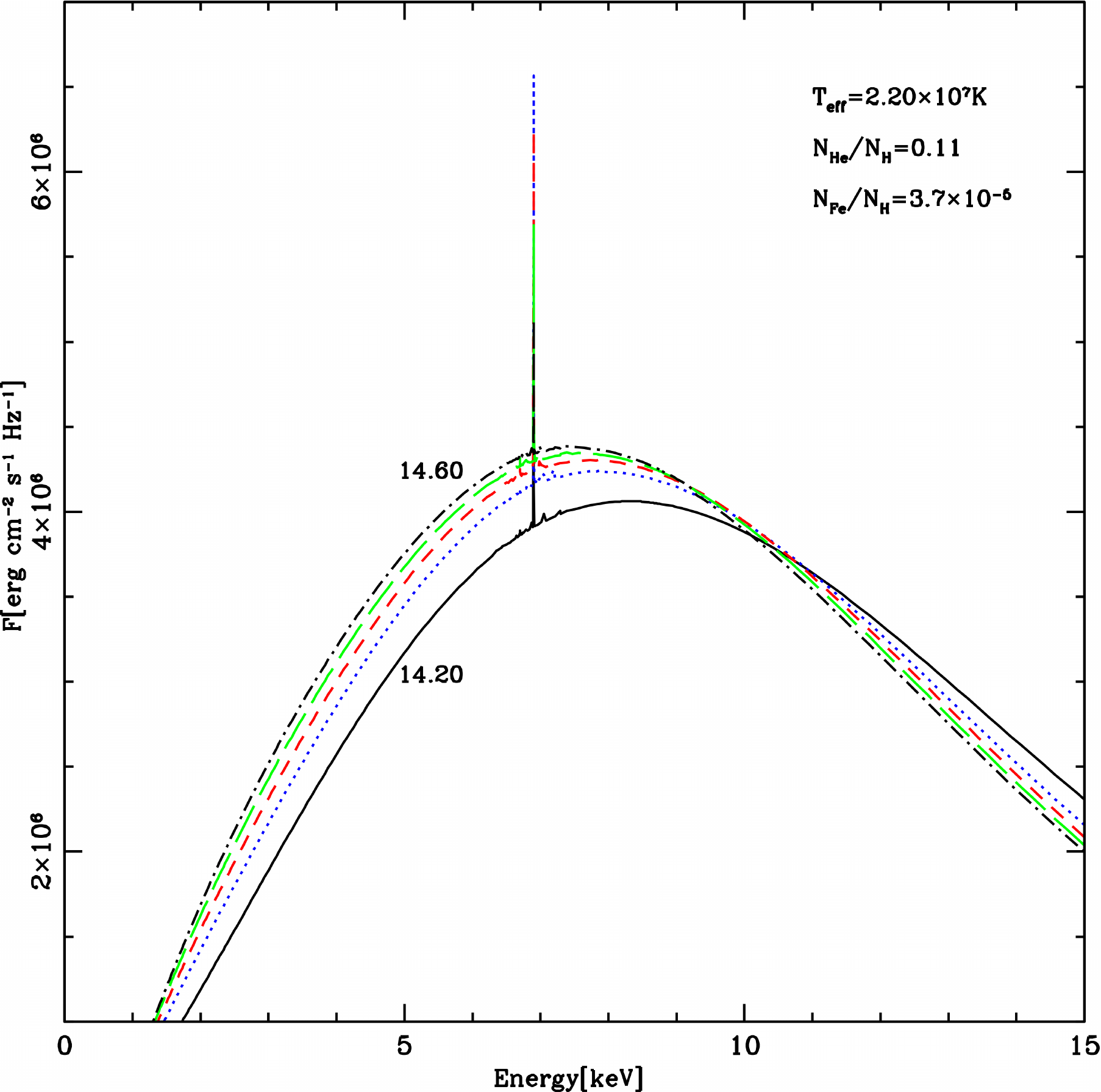}
\includegraphics[scale=0.46]{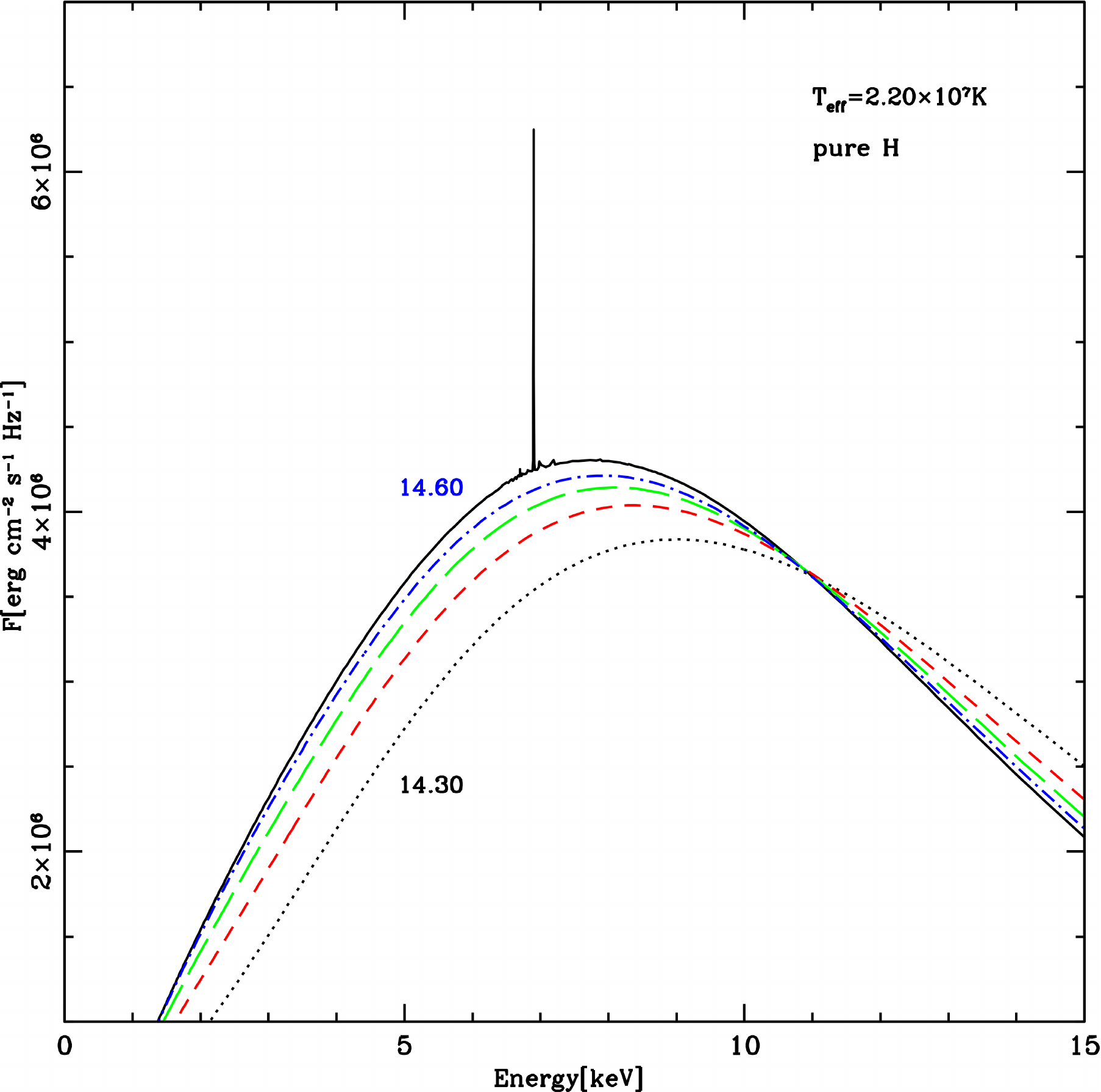}
\end{center}
\caption{Theoretical local spectrum of hot neutron star atmosphere with parameters: $T_{\rm
eff}=2.20\times 10^7$ K and different surface gravities: left panel from $\log(g)=14.20$ 
to 14.60, whereas right one from 14.30 to 14.60. The assumed hydrogen-helium-iron composition if of 
solar proportion (left panel) and pure hydrogen atmosphere (right panel). On the right panel 
picture, we add to comparison spectrum of the atmosphere with iron and $\log(g)=14.40$ (solid, black 
line).}
\label{atm_widmo}
\end{figure}

We solve the model atmosphere assuming constrains of hydrostatic and radiative equilibrium.
We are aware that for atmosphere in motion this assumption is too strong, however, such
models were widely used to fit the X-ray spectra (see e.g. \cite{suleimanov17,medin16}). The
influences of the magnetic field and accretion onto the neutron star are not included. Our code 
takes into account energy-dependent opacities of hydrogen, helium and heavy element ions in LTE.
The ionization equilibrium is fully solved, allowing for the appearance of iron lines for specific 
initial parameters \cite{majczyna05b}. We neglect the effects of electron degeneracy, which are 
unimportant in the hot atmospheres relevant to our studies. Examples of the theoretical local 
spectra for one value of effective temperature and several surface gravities are shown in the 
Figure \ref{atm_widmo}. Near the maximum flux, a few emission iron lines are clearly seen.

Right panel of Figure \ref{atm_widmo} shown spectra of hot neutron star calculated by ATM24 code for 
the same effective temperature $T_{\rm eff}=2.20\times 10^7\,$K and logarithm of surface gravity 
from $\log(g)=14.30$ up to $\log(g)=14.60$. We assumed pure hydrogen atmosphere. For comparison we 
also add spectrum of the atmosphere composed by mixture of hydrogen, helium and iron in following 
proportions: $N_{He}/N_H=0.11$ and $N_{Fe}/N_H=3.7\times 10^{-5}$ and $\log(g)=14.40$.

\section{SIMULATED SPECTRUM}
\label{fake}

ATHENA is a future mission; therefore for the aim of this paper, we simulated a spectrum
which will be detected by the WFI instrument. We used publicly available calibration
files\footnote{http://www.mpe.mpg.de/ATHENA-WFI/response\_matrices.html} provided by the ATHENA 
mission  team. The effective area at 1~keV is 1.4 m$^2$. To simulate the observed spectrum with 
WFI detector, we used ``fake'' command in {\tt xspec 12.6.0} fitting package \citep{xspec}.
The above command works on theoretical model and simulates the data taking into account 
WFI/ATHENA responces and background files for the newest design of mirror with 15 rows$^1$. 
The obtained data file is accompanied by relevant simulated new background file.  In the
case of simulated spectrum and background all errors are Poissonian.

\begin{table*}[!h]
\caption{ Fake spectra parameters for A, B, and C model atmospheres: hydrogen column density 
$N_H$, effective temperature $T_{\rm eff}$, surface gravity $\log g$, gra\-vitational redshift $z$ 
and normalization factor $N_{\rm ATM}$. In last three rows we display corresponding masses, radii 
and fluxes, $F$, for those sources.}
\label{tab:fejki}
\begin{center}
\begin{tabular}{|c|c|c|c|}
\hline
name                     & A & B & C \\ \hline 
$N_H$ [cm$^{-2}$]  & $0.8\times 10^{22}$ & $0.8\times 10^{22}$ & $0.8\times 10^{22}$  \\ \hline
$T_{\rm eff}$  [K] & $2.19 \times 10^7$  & $2.20 \times 10^7$ & $2.21 \times 10^7$ \\ \hline
$\log g$ [cgs]        & 14.25   & 14.30   & 14.35        \\ \hline 
$z$                       & 0.240   & 0.300   & 0.350        \\ \hline
$N_{\rm ATM}$   &2.5$\times10^{-24}$&2.5$\times10^{-24}$ & 2.5$\times10^{-24}$ \\ \hline
$M\,$[M$_\odot$] & 1.297   & 1.653   & 1.869        \\ \hline
$R\,$[km]             & 10.956 & 11.954 & 12.230      \\ \hline
$F$ [erg~cm$^{-2}$~s$^{-1}$] & 4.51 $\times 10^{-10}$ & 4.42 $\times 10^{-10}$ & 4.32  $\times 10^{-10}$ \\
\hline
\end{tabular}
\end{center}
\end{table*}

To produce simulated data, we chose three models with various para\-meters for the neutron star 
atmosphere: effective temperature  $T_{\rm eff}$, surface gravity $\log g$, 
gravitational redshift $z$ and normalization factor $N_{\rm ATM}$ given in Table~\ref{tab:fejki}. 
We name those models as A, B, C respectively. 
The values of $\log g$ and $z$ correspond to particular masses and radii of a neutron stars 
also given in the above Table (see Section~\ref{results} for relation between parameters).
The normalization factor is directly related with the ratio of the neutron star radius to the 
distance $D$ as $(R/D)^2$. We normalize our models in such a way, that  the values of observed 
fluxes 
correspond to semi-bright Galactic X-ray sources. All observed fluxes are given in the last row 
of Table~\ref{tab:fejki}.  

These parameters are not related to any  particular existing neutron star, but 
compact objects with these parameters certainly could be realized in  nature. 
The chemical composition was assumed as a mixture of hydrogen $N_{\rm He}/N_{\rm H}=0.11$ and iron  
$N_{\rm Fe}/N_{\rm H}=3.7\times 10^{-5}$ (number abundances). Corresponding relative mass abundances 
are: $M_{\rm H}$=0.6950, helium $M_{\rm He}$=0.3035
and iron $M_{\rm Fe}$=1.425$\times 10^{-3}$. Finally all our models were multiplied by interstellar 
absorption model ({\sc tbabs} in {\tt xspec}) with  the same assumed hydrogen column density 
$N_{\rm H}=0.80\times 10^{22}\,$cm$^{-2}$. We set the time exposure $t_{\rm exp}$ equal to 1 
second. Such a value of $t_{\rm exp}$  can be used for objects like isolated neutron stars or X-ray 
transients in the period when the neutron star is not accreting matter.

\begin{figure}[!h]
\begin{center}
\includegraphics[scale=0.5,angle=270]{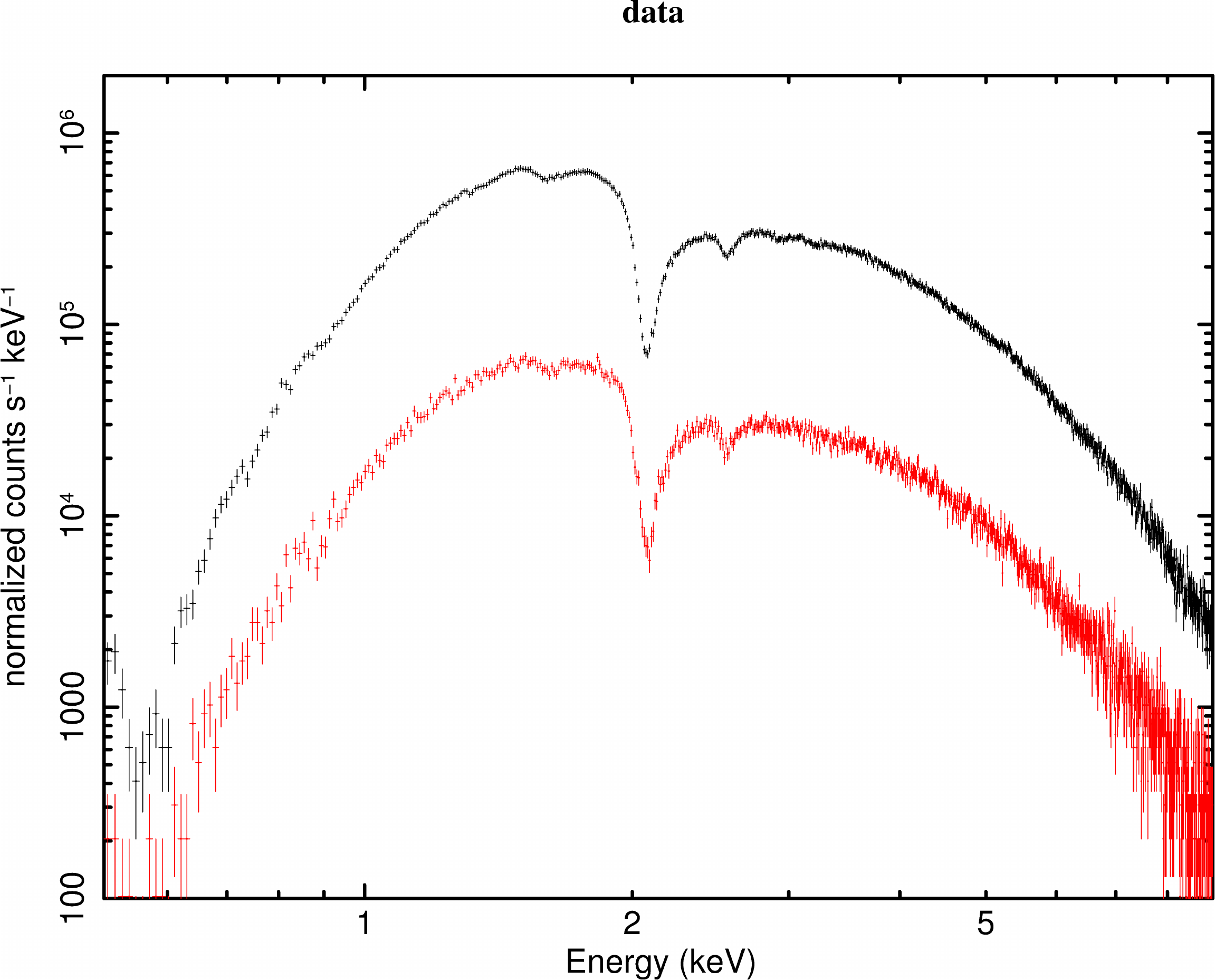}
\end{center}
\caption{Simulated WFI spectra for arbitrarily chosen parameters: $T_{\rm eff}=2.2\times
10^7\,$K, $\log(g)=14.3$, $z=0.30$ and $N_{\rm ATM}=2.5\times 10^{-24}$ (fake spectrum B; black 
crosses) and $N_{\rm ATM}=2.5\times 10^{-25}$ (red). Systematic errors on the level of 3\% are 
taken into account. The chemical composition is assumed as mixture of hydrogen, helium and iron in 
the solar proportion. The time exposure $t_{\rm exp}$ is equal to 1 second. The overall data are 
shaped by ATHENA mirrors and WFI effective area.}
\label{fejk_atena}
\end{figure}

In the case of X-ray bursters, however, the situation is more complicated. In such objects the 
method based on fitting observed spectra  should be used for photospheric radius expansion 
bursts in the touchdown phase in the hard state. In such a situation the exposure time should 
be much shorter of the order of tenths of a second. Therefore only very sensitive detectors 
with large effective area and time resolution of the order of microseconds can be used. In the 
past, the best observational spectra were provided by RXTE which allowed collection of many 
counts during 0.1 second. CCD type detectors used onboard of Chandra and XMM-Newton 
missions are not fast enough to collect a sufficient amount of photons even during maximum 
burst phase. Therefore, the fitting procedure of continuum emission for mass and radius 
determination would not be very precise for data from those satellites. Besides the case of X-
ray bursters presented in this paper, our method could also be used for isolated neutron stars 
or transient objects. However, for different sets of physical parameters, new model 
computations would be required.

Since our analysis relies on X-ray spectra, systematic errors can be important 
\citep{arnaud_book2011,lee2011,xu2014}. In order to put constrains on the parameters from the 
spectral shape (as in the  case of our paper), only relative area on-axis systematic errors are 
important, which influence the observed spectral shape, and therefore estimation of model parameters 
(in our case neutron star mass and radius). Those errors depend on the detector calibration and for 
ATHENA mission the expected value is on the level of 3\% (ATHENA Calibration requirement document, 
ESA Technical Note). Therefore, after our fake spectra have been made, we added systematic errors 
on the level of 3\% using {\tt xspec} ftool {\sc grppha}. 

Figure \ref{fejk_atena} shows two simulated spectra for parameters of model B and two different 
assumed  unabsorbed fluxes - $f_{2-10keV}=4.42\times 10^{-10}\,$erg$\,$cm$^{-2}\,$s$^{-1}$ (black 
crosses) and $f_{2-10keV}=4.41\times 10^{-11}\,$erg$\,$cm$^{-2}\,$s$^{-1}$ (red). In case of 
spectrum with larger flux, for a 1 second exposure time, we have collected 2.36$\times 10^6$ 
photons, enough to obtain our science goal. 

\section{FITTING PROCEDURES}
\label{fitting}

Our method of determination of neutron star parameters is based on the fitting of theoretical
spectra to the observed one -- in this case to the WFI/ATHENA fake spectrum. We used the 
fake data with higher unabsorbed flux (black crosses at Fig.~\ref{fejk_atena}) for further analysis.

\begin{figure}[]
\begin{center}
\includegraphics[scale=0.5,angle=270]{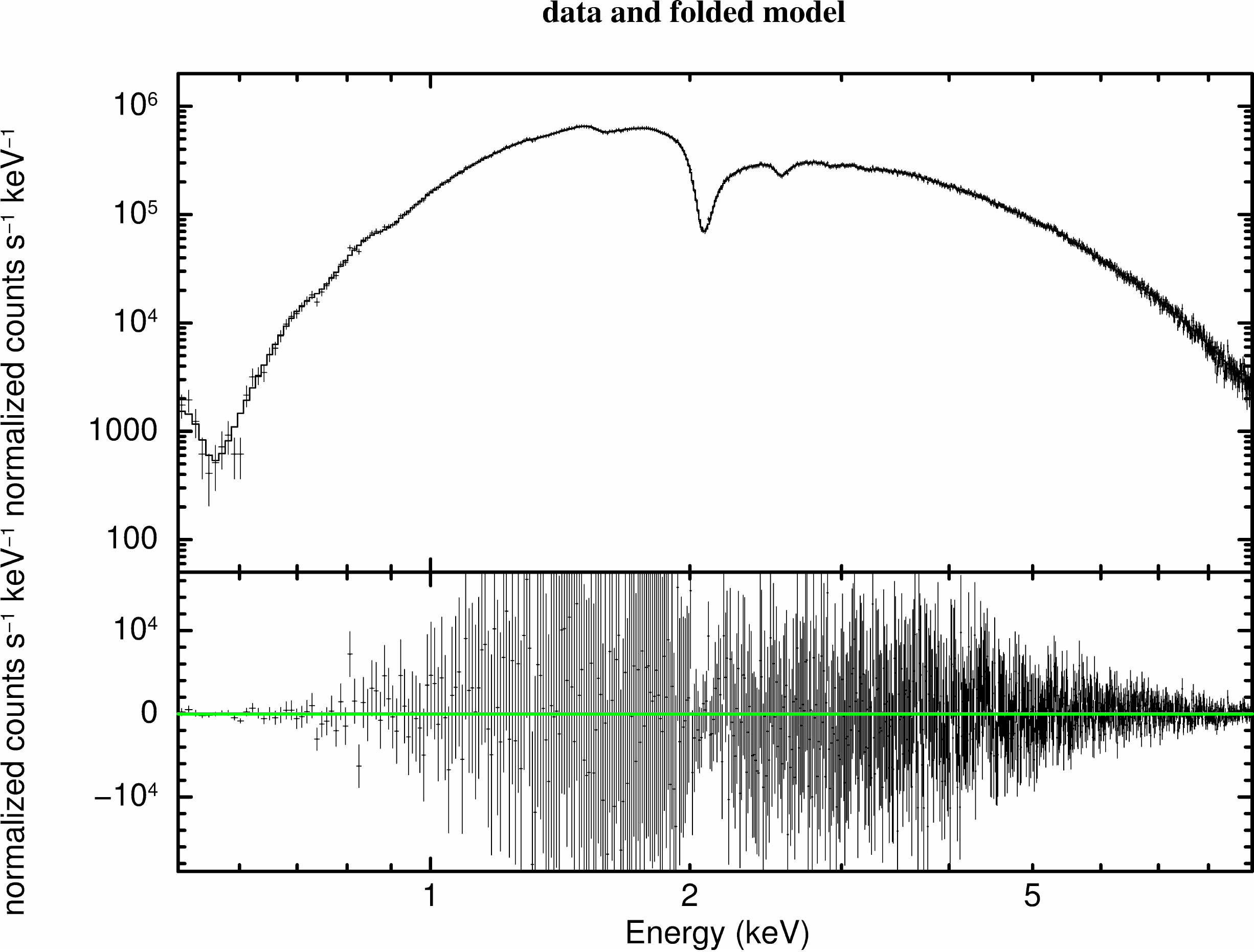}
\end{center}
\caption{Top panel: data (fake spectrum B) and our best fit model, while bottom panel shows residua (data minus the 
folded model). Parameters of the fit are as follow: $T_{\rm eff}=2.18\times 10^7\,$K, log(g)=14.28, 
z=0.305, $N_{ATM}=2.56\times 10^{-24}$ and $N_{\rm H}=0.804\times 10^{22}\,$cm$^{-2}$ and $\chi^2$=518.52/869.}
\label{fit_best}
\end{figure}

The theoretical models used do fit the fake spectrum
are constructed for one chemical composition, given above. Four parameters as: effective
temperature, surface gravity, gravitational redshift and normalization are free parameters in our
fitting procedure. In addition, $T_{\rm eff}$ and $\log g$ are input parameters in our atmospheric
ATM24 numerical simulations. 
We calculated an extensive grid of theoretical spectra (nearly 5000 models) with the chemical
composition given above. In our initial grid of models, the effective temperature ranges from 
$10^7$~K to $2.70\times 10^7\,$K with step $\Delta T_{\rm eff}=0.02\times 10^7\,$K, surface 
gravity $\log g$ from the critical gravity up to 15.0 (cgs) with $\Delta \log g=0.02$.
However, we found that the error of $\log g$ is smaller than $\Delta \log=0.02$, so it was obvious 
that we needed a denser grid of models. Thus, we have chosen smaller steps of parameters around our 
reference values for the effective temperature and gravity. For the effective temperatures in the 
range from $T_{\rm eff}=2.18\times 10^7\,$K to $2.22\times 10^7\,$K steps were $\Delta T_{\rm 
eff}=0.01\times 10^7\,$K, and for surface gravity ranging from $\log g =13.9$ to $14.60$ we have 
chosen $\Delta \log g=0.01$. All our models were converted to FITS format \citep{fits}, suitable to 
{\tt xspec 12.6.0} package \citep{xspec}. The latter software was used also to fit our models to 
simulated WFI/ATHENA spectrum.

For each given combination of values of $T_{\rm {eff}}$ and $\log(g)$, the surface redshift 
$z$ was varied from 0.1 to 0.6 with steps of 0.005. The value of model normalization factor $N_{\rm 
ATM}$ corresponding to the best fit was determined. During the fitting procedure the value of 
hydrogen column density ($N_H$) in model of Galactic absorption ({\sc tbabs} model in {\tt xspec}) 
was a free parameter. Therefore, we obtained a large,  5-dimensional 
table of $\chi^2$ for one assumed chemical composition. Then, from this huge table, we extracted 
one set of four parameters ($T_{\rm eff}$, $\log{(g)}$, $z$ and $N_{\rm ATM}$) corresponding 
to the fit with the lowest value of $\chi^2$. We found also 1, 2 and 3$\sigma$ confidence
levels in $\log(g)-z$ parameters space, requiring that $\chi^2_{\rm min} <  \chi^2 < \chi^2_{\rm 
min}+\Delta\chi^2$, and additionally that $0.1 < M < 3\,$M$_\odot$. The value of $\Delta \chi^2$ 
corresponds to the 1, 2 and 3$\sigma$ confidence levels for two free pa\-ra\-me\-ters 
\citep{numerical}. The best fitted model and residua are presented in Figure~\ref{fit_best}.

\begin{table*}[!h]
\caption{ Best fit parameters and 1$\sigma$ and 2$\sigma$ errors (in parenthesis).}
\label{tab:wyniki}
\begin{center}
\begin{tabular}{|c|c|c|c|c|c|c|c|c|c|}
\hline
\multicolumn{10}{|c|}{fake spectrum A} \\ \hline
\multicolumn{2}{|c|}{$N_H$}& \multicolumn{2}{|c|}{log(g)} & \multicolumn{2}{|c|}{z} &\multicolumn{2}{|c|}{M[M$_\odot$]} & \multicolumn{2}{|c|}{R[km]} \\ \hline
0.802 &$^{+0.002}_{-0.002}(^{0.004}_{0.004})$& 14.26 & $^{+0.03}_{-0.02}$($^{+0.03}_{-0.03}$) & 0.255 &$^{+0.005}_{-0.020}$($^{+0.020}_{-0.025}$)& 1.40&$^{+0.05}_{-0.15}$($^{+0.20}_{-0.24}$)& 11.32&$^{+0.32}_{-0.94}$ ($^{+1.24}_{-1.32}$)\\ \hline
\multicolumn{10}{|c|}{fake spectrum B} \\ \hline
0.804 & $^{+0.002}_{-0.002}(^{0.004}_{0.004})$&14.28 &$^{+0.04}_{-0.0005}(^{+0.06}_{-0.01})$  & 0.305 &$^{+0.015}_{-0.010}(^{+0.025}_{-0.015})$& 1.78&$^{+0.05}_{-0.13}(^{+0.18}_{-0.25})$ &12.71 & $^{+0.19}_{-0.95}(^{+0.87}_{-1.64})$ \\ \hline
\multicolumn{10}{|c|}{fake spectrum C} \\ \hline
0.800 & $^{+0.002}_{-0.002}(^{0.004}_{0.003})$ & 14.32 &$^{+0.04}_{-0.02}(^{+0.06}_{-0.03})$ & 0.360 &$^{+0.015}_{-0.015}(^{+0.030}_{-0.025})$& 2.09&$^{+0.15}_{-0.22}(^{+0.24}_{-0.31})$&13.44 & $^{+0.99}_{-1.08}(^{+1.32}_{-1.88})$ \\ \hline
\end{tabular}
\end{center}
\end{table*}

\begin{table*}[!h]
\caption{ Best fitted parameters for our three fake spectra, A, B and C. Confidence values for 
1$\sigma$, 2$\sigma$ and 3$\sigma$ are given in next three rows for each model.}
\label{tab:kontury}
\begin{center}
\begin{tabular}{|c|c|c|c|c|}
\hline
\multicolumn{5}{|c|}{fake spectrum A} \\ \hline
                 & $z$          & $\log(g)$ [cgs]  & $M$ [M$_\odot$] & $R$ [km]  \\ \hline
best par.          &  0.255      & 14.26                 &  1.399               &  11.315  \\   \hline   
1$\sigma$ & 0.235 -- 0.260 & 14.24 -- 14.29   & 1.252 -- 1.444   & 10.371 -- 11.637 \\   \hline
2$\sigma$ & 0.230 -- 0.275 & 14.23 -- 14.29   & 1.155 -- 1.595   &   9.992 -- 12.556 \\   \hline
3$\sigma$ & 0.220 -- 0.258 & 14.22 -- 14.31   & 1.049 -- 1.711   &   9.392 -- 13.068 \\   \hline
\multicolumn{5}{|c|}{fake spectrum B} \\ \hline
                 & $z$         & $\log(g)$ [cgs]  & $M$ [M$_\odot$] & $R$ [km]  \\ \hline
best par.    & 0.305      & 14.28              & 1.776                  &  12.705 \\   \hline          
1$\sigma$ & 0.295 -- 0.320 & 14.28 -- 14.32   & 1.647 -- 1.825    & 11.758 -- 12.892 \\   \hline
2$\sigma$ & 0.290 -- 0.330 & 14.27 -- 14.34   & 1.531 -- 1.959    & 11.066 -- 13.573 \\   \hline
3$\sigma$ & 0.280 -- 0.335 & 14.26 -- 14.35   & 1.429 -- 2.054    & 10.494 -- 14.083 \\   \hline
\multicolumn{5}{|c|}{fake spectrum C} \\ \hline
                  & $z$          & $\log(g)$ [cgs]  & $M$ [M$_\odot$] & $R$ [km]  \\ \hline
best par.    & 0.360        & 14.32              &  2.090                  &  13.437 \\   \hline          
1$\sigma$ & 0.345 -- 0.375 & 14.30-- 14.36   & 1.869 -- 2.235    & 12.230 -- 14.243 \\   \hline
2$\sigma$ & 0.335 -- 0.390 & 14.29 -- 14.38  & 1.782 -- 2.334    & 11.558 -- 14.752 \\   \hline
3$\sigma$ & 0.325 -- 0.395 & 14.28 -- 14.40  & 1.630 -- 2.486    & 10.761 -- 15.456 \\   \hline
\end{tabular}
\end{center}
\end{table*}

\section{RESULTS}
\label{results}

As a result of fitting the simulated WFI/ATHENA data we determined the effective temperature
$T_{\rm eff}$, surface gravity $\log g$ and gravitational redshift $z$. The two latter parameters 
are converted into mass and radius of the neutron star following \citet{majczyna05}:
\begin{equation}
R=\frac{zc^2}{2g}\frac{(2+z)}{(1+z)},
\end{equation}
and
\begin{equation}
M=\frac{z^2c^4}{4gG}\frac{(2+z)^2}{(1+z)^3},
\end{equation}
where $G$ is the gravitational constant and $c$ is the speed of light.

\begin{figure}[!h]
\begin{center}
\includegraphics[scale=0.3]{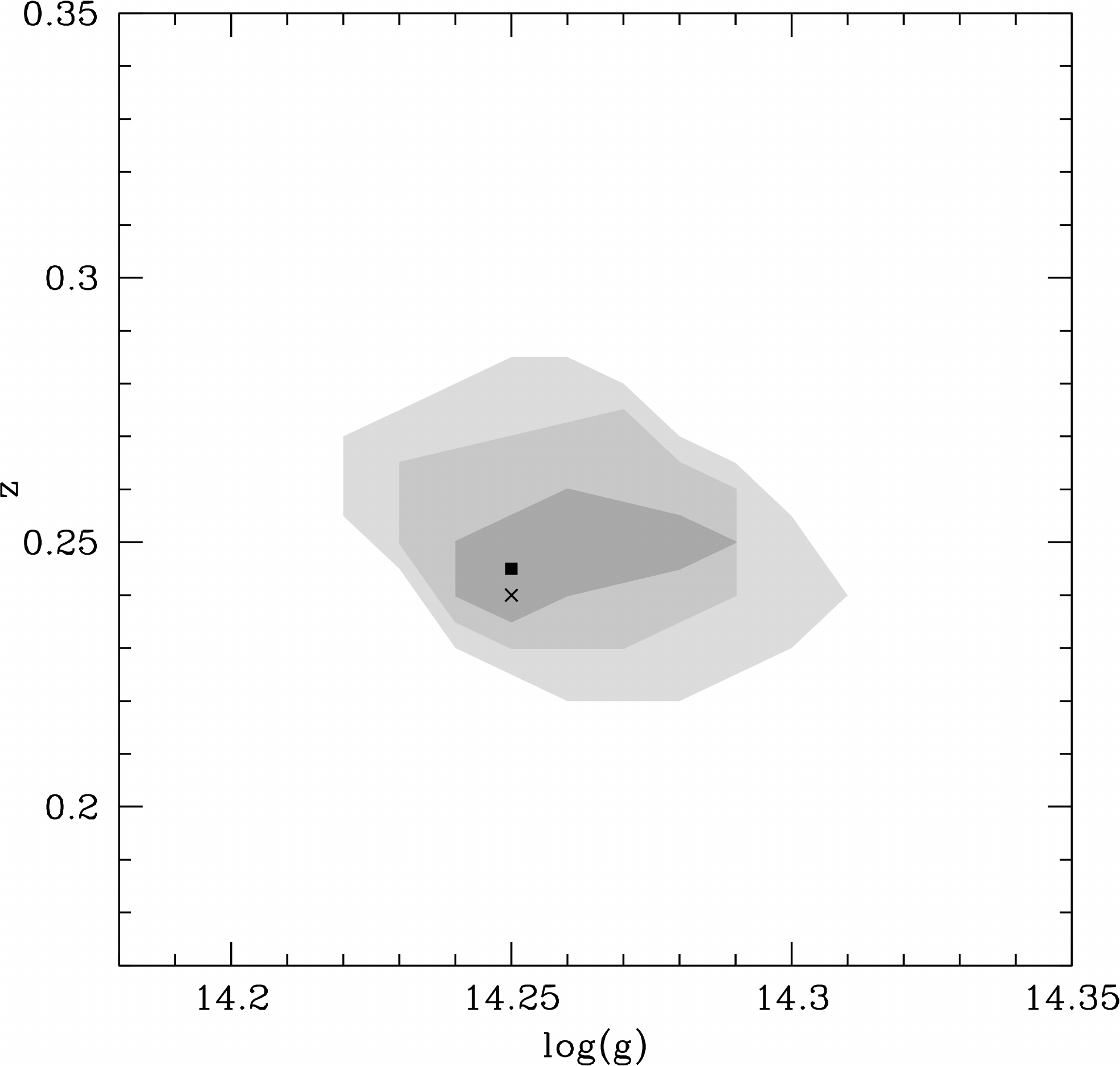}
\includegraphics[scale=0.3]{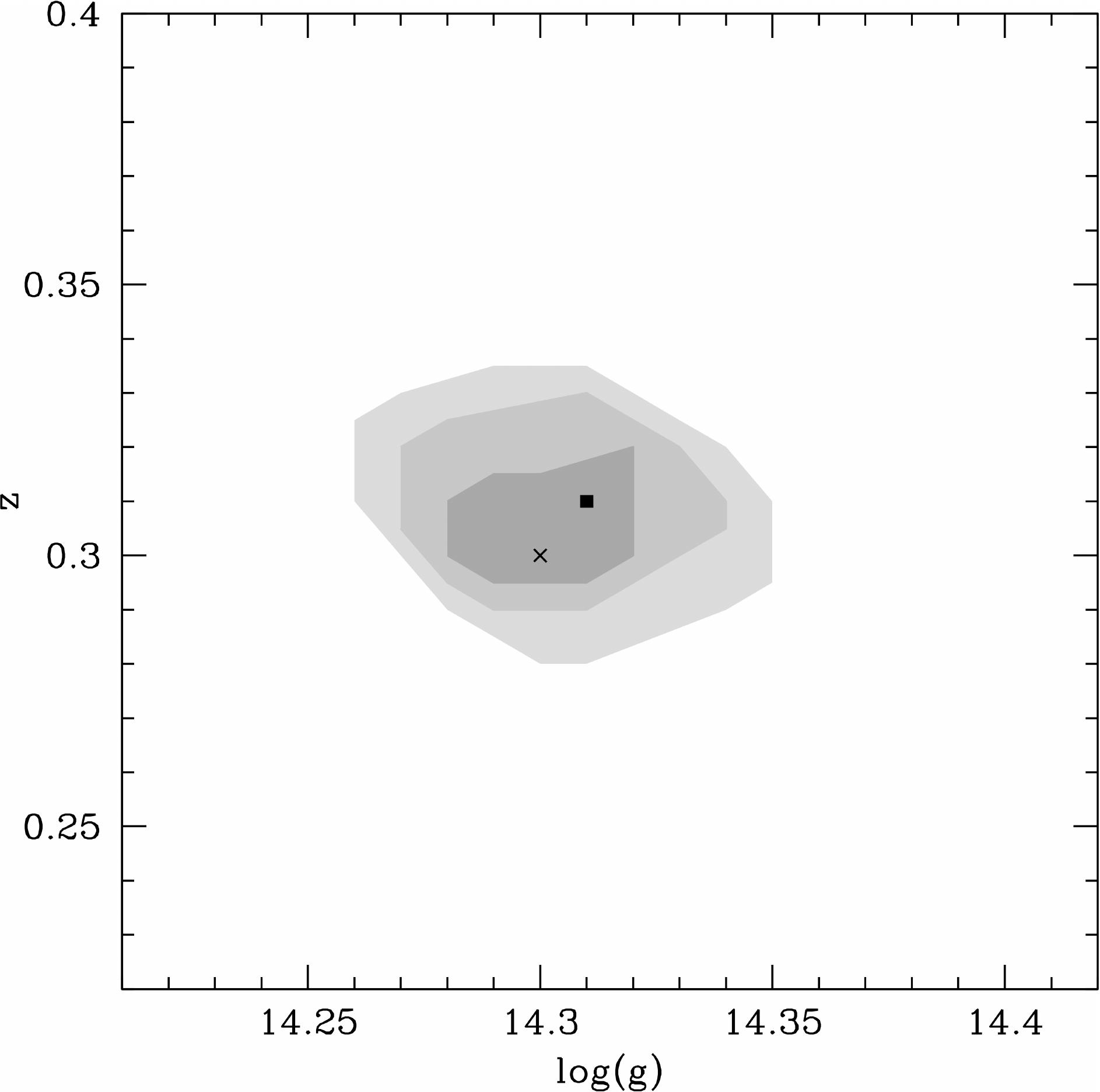}
\includegraphics[scale=0.3]{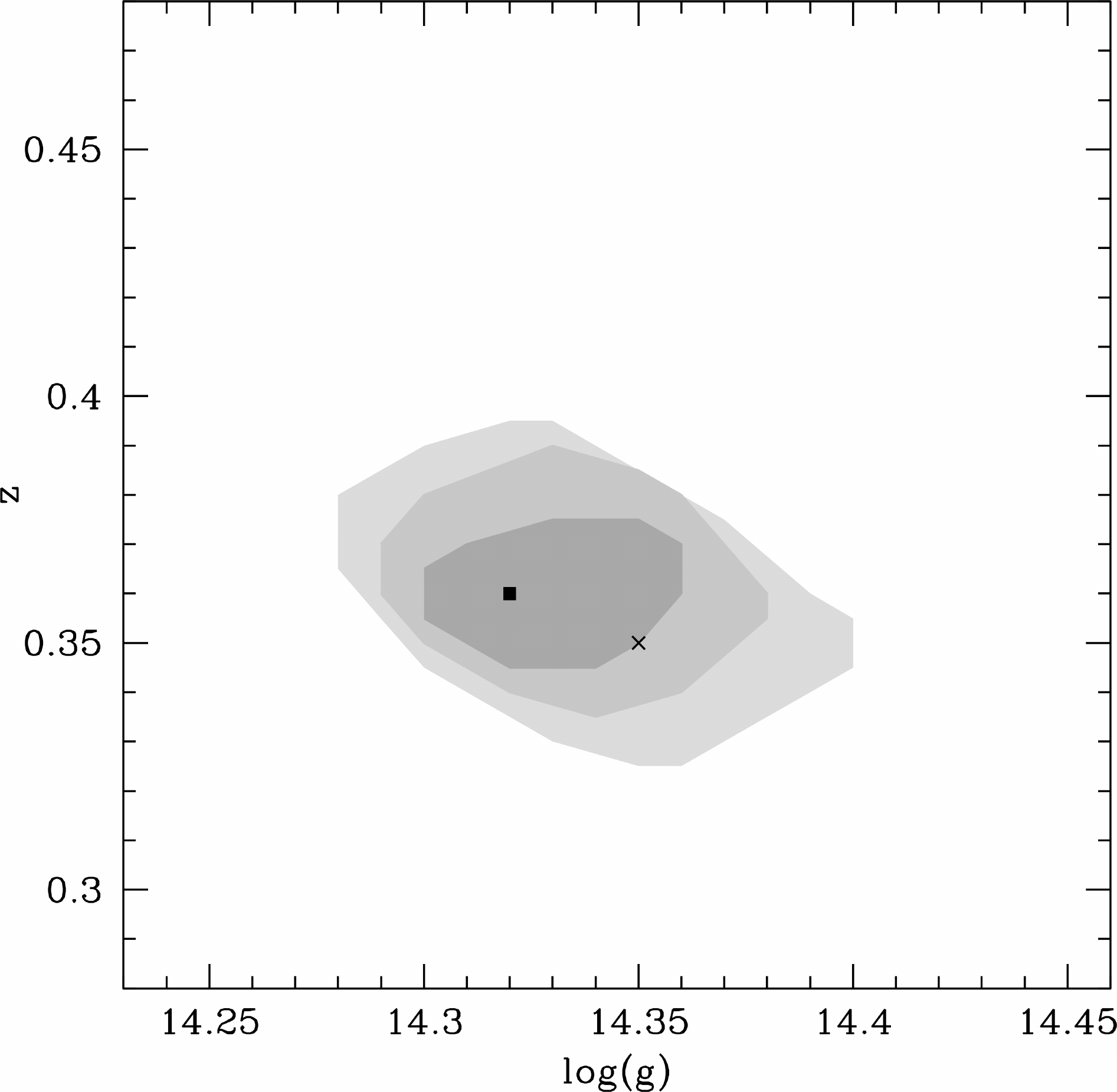}
\end{center}
\caption{ 1, 2 and 3$\sigma$ confidence contours for two free parameters: redshift and surface
gravity for three models A -- left panel, B -- middle panel, and C -- right panel. Black 
cross denotes our reference values, while black dot is the best fitted value.}
\label{kontury}
\end{figure}

\begin{figure}[!h]
\begin{center}
\includegraphics[scale=0.46]{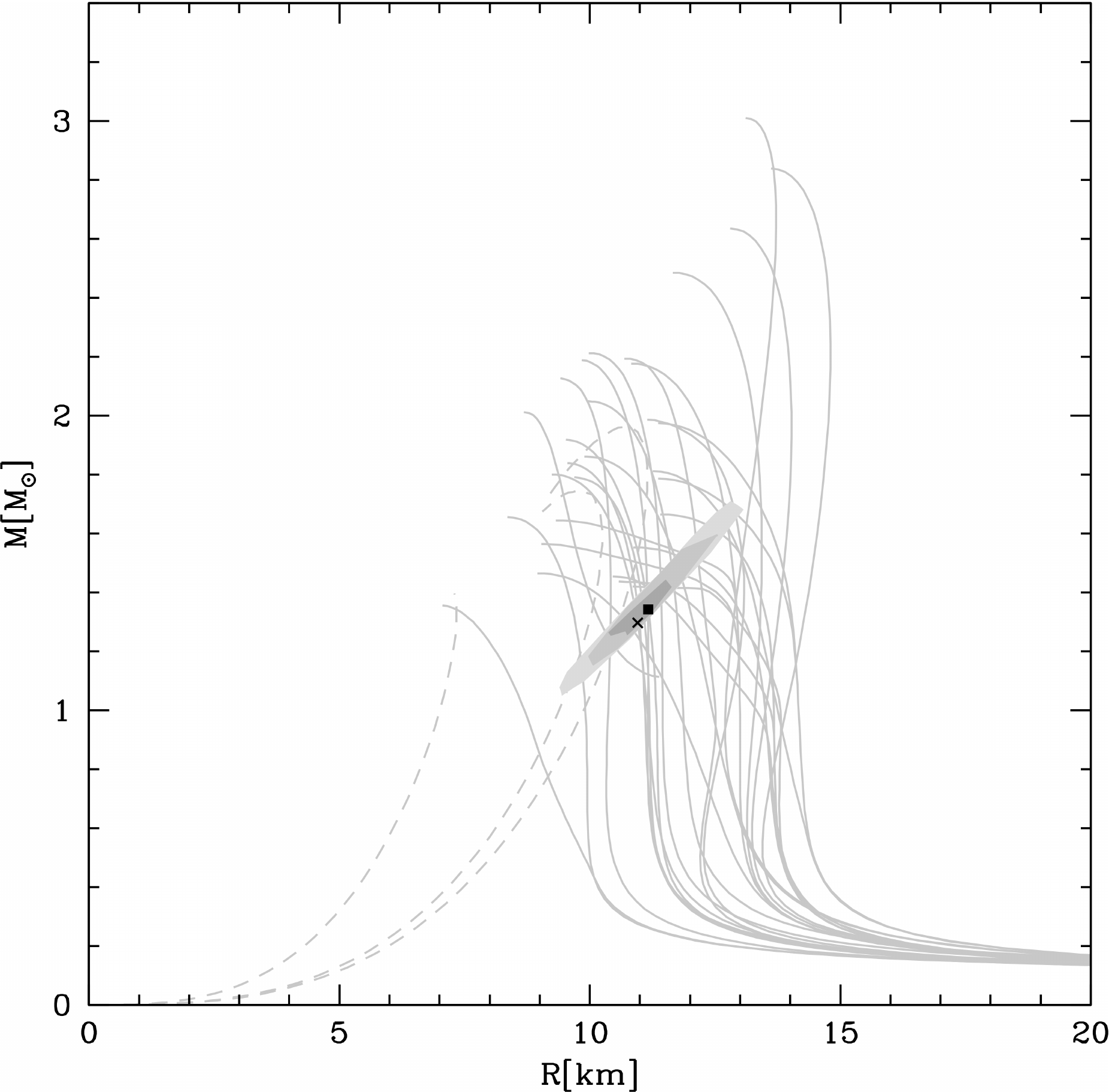}
\includegraphics[scale=0.46]{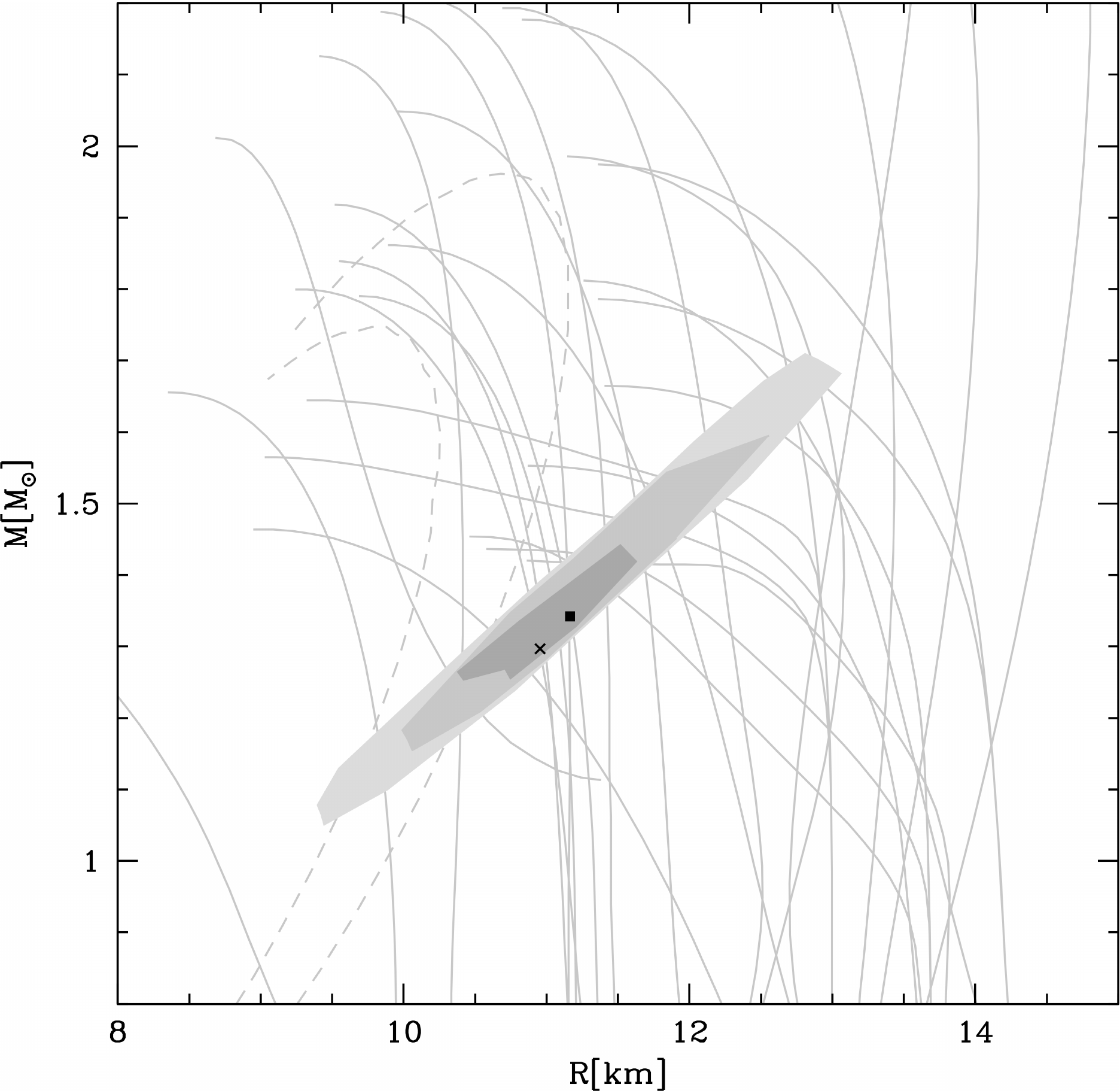}
\end{center}
\caption{ 1, 2 and 3$\sigma$ confidence contours for two free parameters: mass and radius for 
model A. Right panel is  the enlarged version of the left panel. Black point denotes our best fit 
mass and radius values $M=1.399\,$M$_\odot$ and $R=11.315\,$km, whereas black cross denotes our 
reference values. Thin gray lines represent possible EOS solutions \citep{haensel07}}
\label{kontury_A}
\end{figure}

Table \ref{tab:wyniki} contains our best fit parameters and the accuracy of their determination 
obtained by using of our method.
Errors were defined as  1$\sigma$ and 2$\sigma$ standard deviations (in parenthesis).
The goal of our fitting procedure is to reproduce assumed values of the pa\-ra\-me\-ters for which
the fake spectrum was calculated (see Sec.~\ref{fake}). We obtained best fit parameters that differ 
slightly from the assumed values, but the difference is less than 1$\sigma$ standard deviations. 
Values of 1, 2 and 3$\sigma$ confidence ranges determined for two free parameters, are presented in 
Table \ref{tab:kontury}, where the minimum $\chi^2$ corresponds to parameters of the neutron star 
that differ slightly from the assumed values

\begin{figure}[!h]
\begin{center}
\includegraphics[scale=0.46]{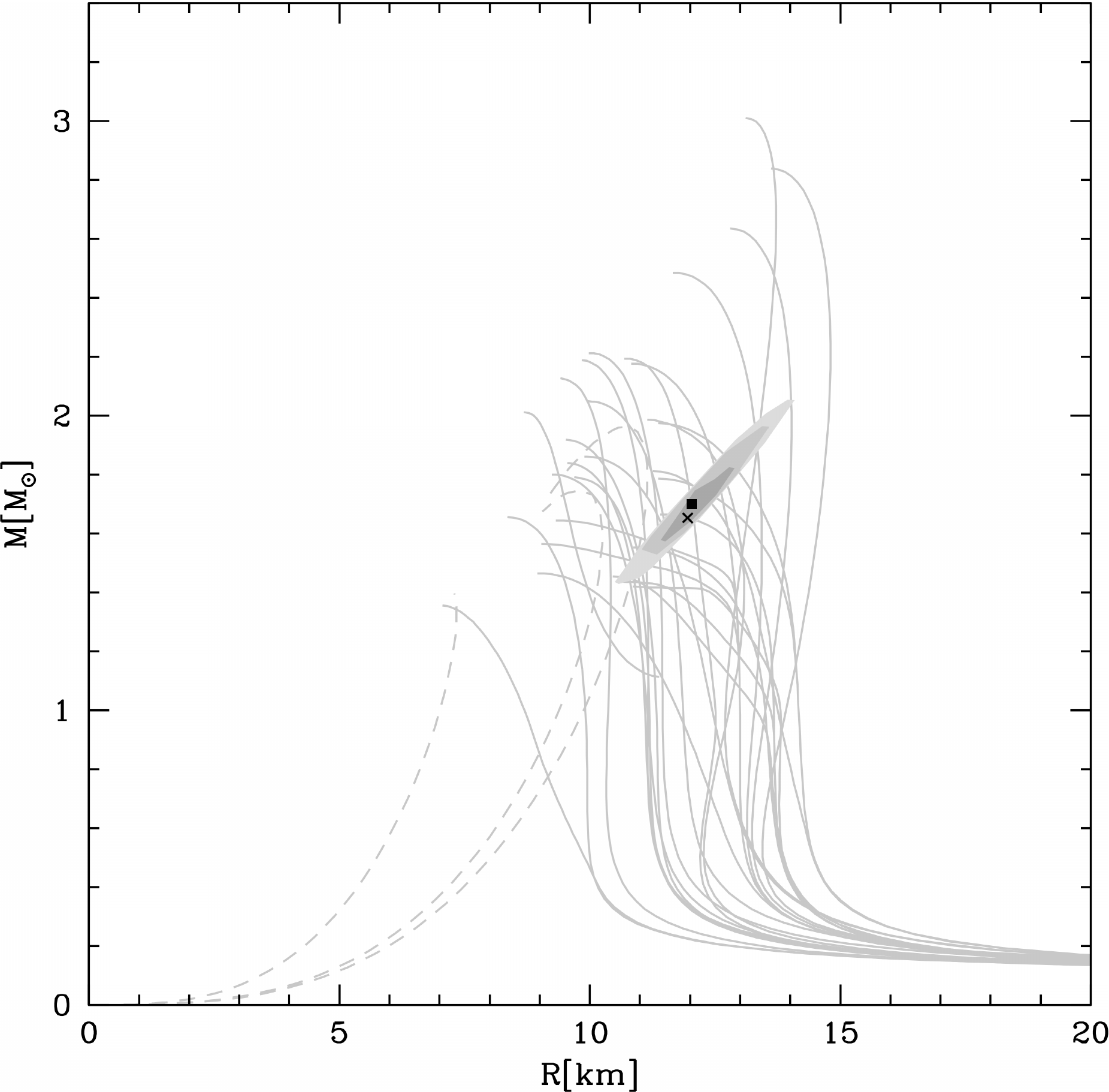}
\includegraphics[scale=0.46]{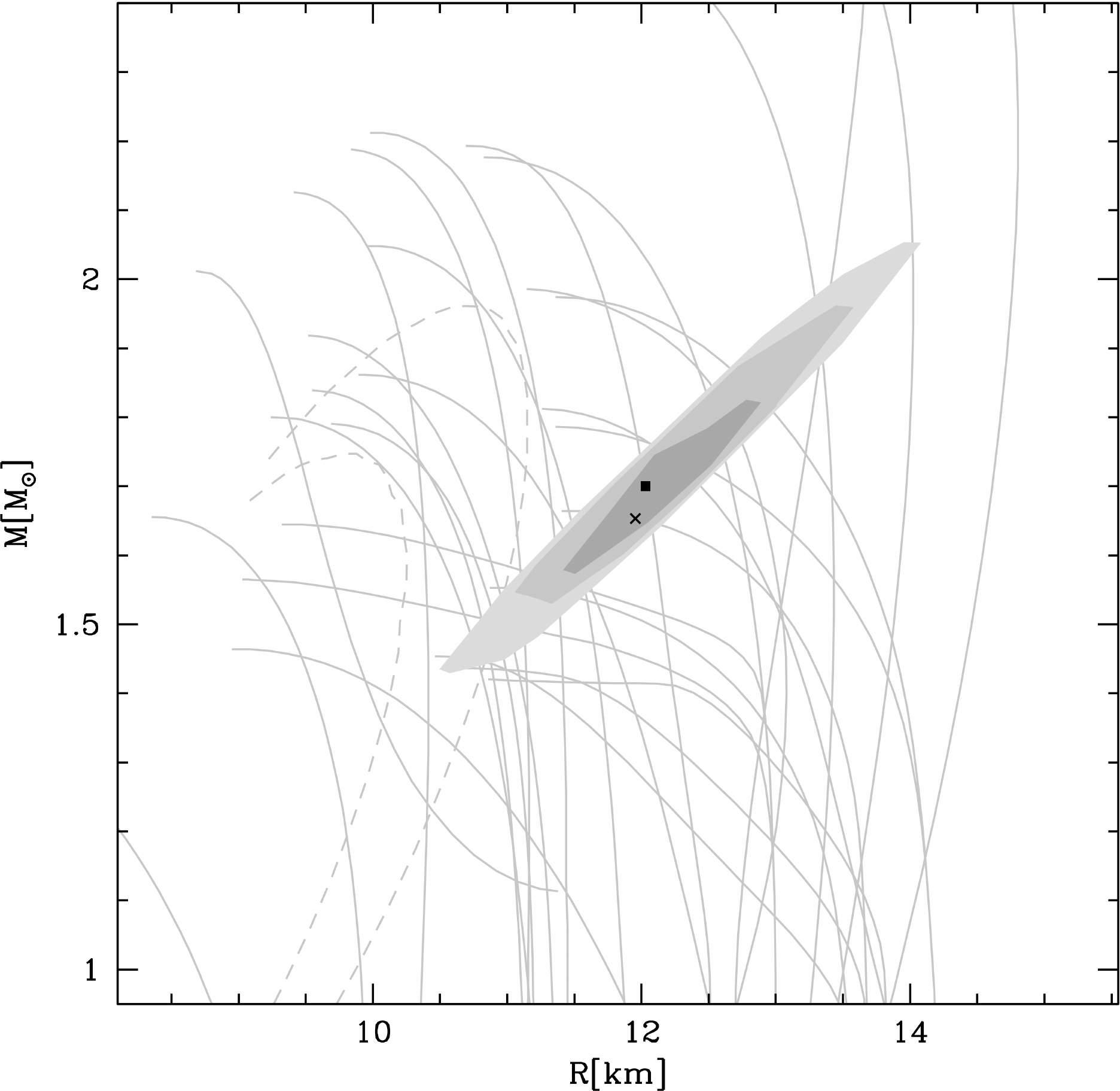}
\end{center}
\caption{ 1, 2 and 3$\sigma$ confidence contours for two free parameters: mass and radius for 
model B. Right panel is  the enlarged version of the left panel. Black point denotes our best fit 
mass and radius values $M=1.776\,$M$_\odot$ and $R=12.705\,$km, whereas black cross denotes our 
reference values. Thin grey lines represent possible EOS solutions \citep{haensel07}}
\label{kontury_B}
\end{figure}

\begin{figure}[!h]
\begin{center}
\includegraphics[scale=0.46]{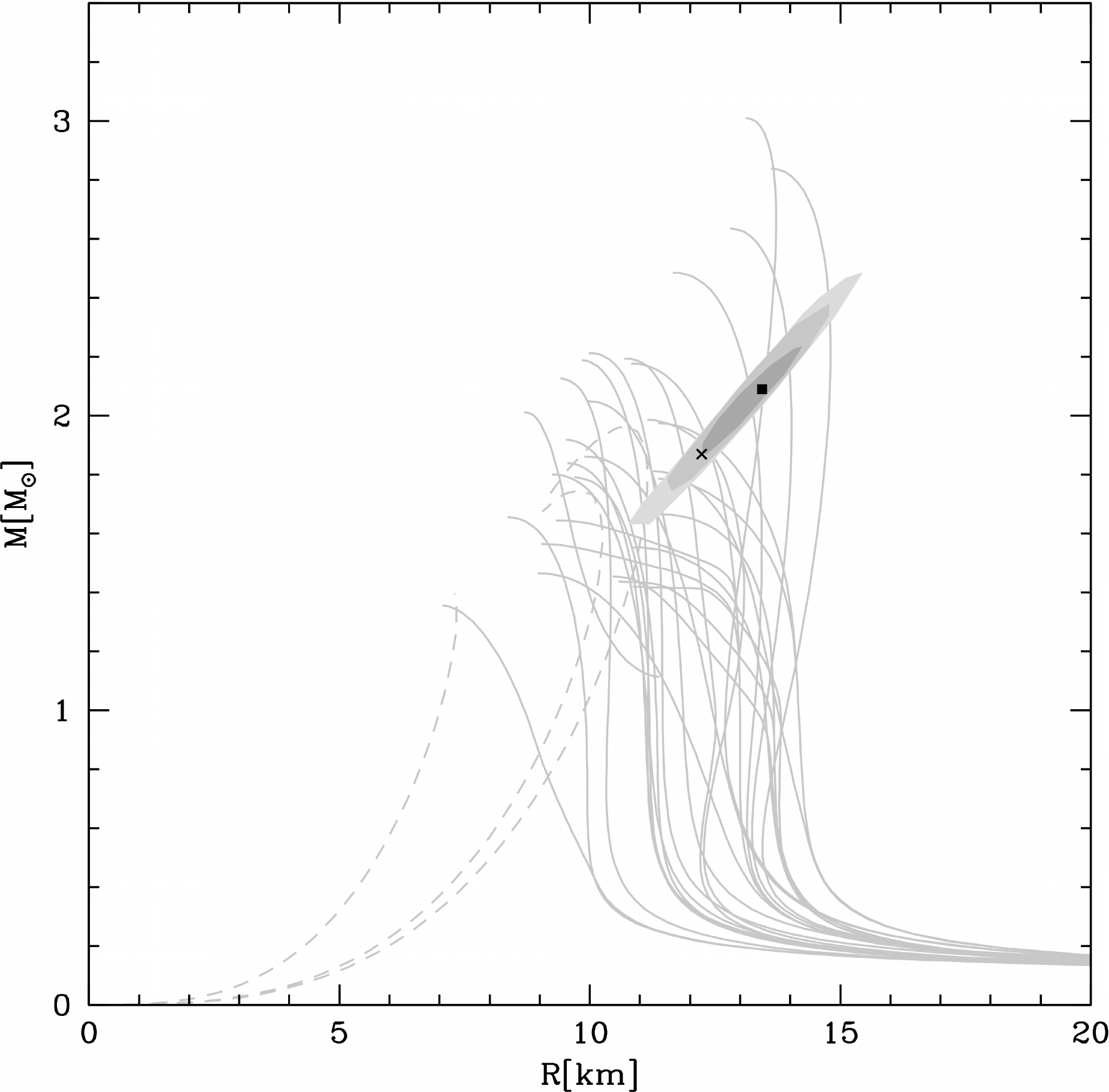}
\includegraphics[scale=0.46]{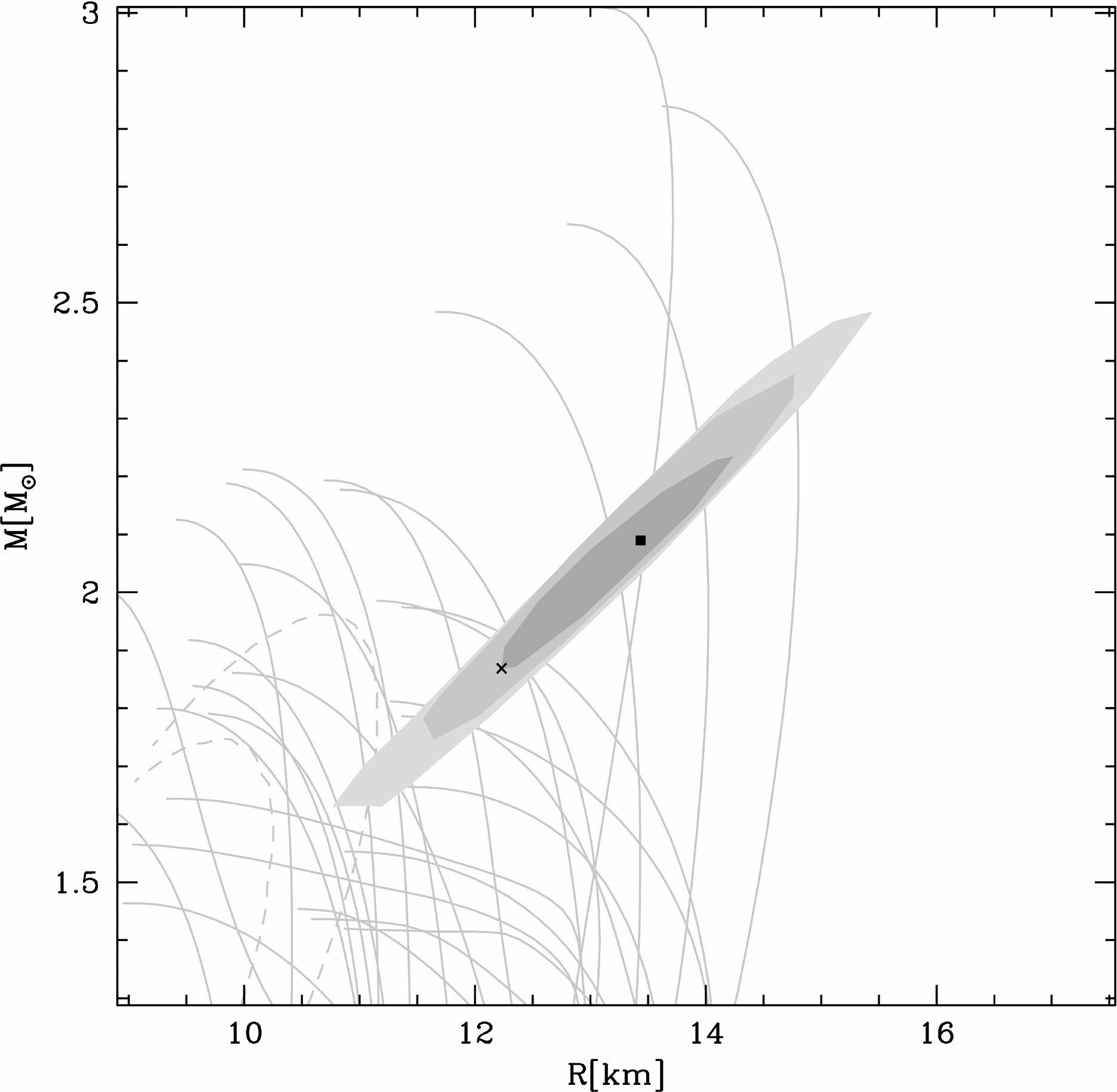}
\end{center}
\caption{ 1, 2 and 3$\sigma$ confidence contours for two free parameters: mass and radius for 
model C. Right panel is  the enlarged version of the left panel. Black point denotes our best fit 
mass and radius values  $M=2.090\,$M$_\odot$ and $R=13.437\,$km, whereas black cross denotes our 
reference values. Thin grey lines represent possible EOS solutions \citep{haensel07}}
\label{kontury_C}
\end{figure}

Figure \ref{kontury} shows the 1,2 and 3$\sigma$ confidence contours,
which are obtained for two free parameters: redshift and surface gravity, for all models. On this 
figure, the black cross denotes the best fitted value of those parameters. The input assumed values 
were denoted as black dot. In all cases, fitted values are within 1$\sigma$ confidence contours.
Corresponding masses and radii for models A, B and C, are presented in Figures~\ref{kontury_A}, 
\ref{kontury_B}, and \ref{kontury_C}, respectively. Furthermore, the grey lines denote possible EOS 
solutions in those figures.

\section{SUMMARY}

Determination of basic parameters of neutron stars is very important for the derivation of the
equation of state of superdense matter. In this paper, we presented the method of mass and radius 
determination for neutron stars. Our method is based on the fitting of theoretical spectra to the 
observed one.  Importantly, our method is independent on the distance, which is proportional to 
the normalization of the model. This is because the normalization factor $N_{\rm ATM }$ is the 
result of our fitting procedure. Therefore, the knowledge of the distance  to the source is not 
necessary if we determined neutron star parameters with our method. Figure \ref{atm_widmo} shows 
our theoretical spectra for two very different chemical compositions.
Shape of the continuum of these spectra is different as well as the location of its maxima.
Those differences indicate, that even tentative knowledge of the chemical composition is crucial
for our method. In case of many neutron stars the chemical composition of their atmospheres is 
known (eg. \citet{goodwin19}).  

We calculated a large grid of theoretical spectra of hot neutron star using ATM24 code. 
assuming effective temperatures $T_{\rm eff}=(1.5-2.7)\times10^7\,$K, logarithm of gravity from 15.0
down to the critical value, and chemical abundances as defined in Section \ref{fitting}.
Parameters of models in the grid changed with steps of $\Delta T_{\rm eff}=0.02\times10^7\,$K and 
$\Delta\log(g)=0.02$; chemical composition was kept the same for all models.

Our goal was to determine the precision of mass and radius determination of the neutron star,
based on the spectra to be obtained with the WFI/ATHENA instrument. Due to lack of real 
ATHENA observations, we simulated three spectra using publicly available WFI calibration
files. We constructed three fake spectra A, B, and C corresponding to three different values of
effective temperatures, surface gravities and redshifts. We have chosen  normalization factors $N_{\rm
ATM}$ which correspond to the observed fluxes of a few hundredths of the Crab 
($\sim 10^{-10}\,$erg$\,$cm$^{-2}\,$s$^{-1}$). ATHENA instrument systematic errors on the 
level of 3\% were taken into account where simulated spectra were created.

Next, we fitted  these fake spectra by a large grid of our theoretical spectra. 
We obtained the best fit ($1\sigma$) for the following parameters of fake spectra:
A $M=1.40^{+0.05}_{-0.15}\,$M$_\odot$ and $R=11.32^{+0.32}_{-0.94}\,$km, 
B $M=1.78^{+0.05}_{-0.13}\,$M$_\odot$ and $R=12.71^{+0.19}_{-0.95}\,$km, 
C $M=2.09^{+0.15}_{-0.22}\,$M$_\odot$ and $R=13.44^{+0.99}_{-1.08}\,$km,
and the corresponding masses and radii for 3$\sigma$ confidence ranges 
$M=1.05-1.71\,$M$_\odot$ and $R=9.38-13.07\,$km,  
$M=1.43-2.05\,$M$_\odot$ and $R=10.49-14.08\,$km,  
$M=1.63-2.49\,$M$_\odot$ and $R=10.76-15.46\,$km, respectively. 

In each above case we determined precision of the mass and radius measurement with errors in the 
range 3--10\% for mass and in the range 2--8\% for radius within the one
sigma confidence error.  All errors ($1\sigma$) are relatively small 
for the WFI/ATHENA detector. We note that the errors defined by 2$\sigma$ confidence ranges
are in the range 11--17\%. Therefore, we showed that our method allows one to constrain the equation of state of 
dense matter which comprises neutron stars using future observations of ATHENA mission. 

\acknowledgements
Special thanks go to Alex Markowitz for helpful discussion and editorial corrections, and to Jan-Willem der Herder and Jorn Wilms for the discussion on systematic errors in ATHENA mission. 
This work was supported by grants 2015/17/B/ST9/03422 and 2015/18/M/ST9/00541 from the Polish National Science Center.

\bibliographystyle{aasjournal}
\bibliography{ns1}
\end{document}